\def\Aw8   {$A_{\omega}^{\delta=0.8}$\ }

\def\n {{$n$~}}
\def\w {{$w$~}}
\def\wn {{$(w-n)$~}}

\def\ie {{\it i.e.}\ }
\def\etal {{\it et al.}}

\def\ksM{km\thinspace s$^{-1}$\thinspace Mpc$^{-1}$}
\def\Ly{Ly$\alpha$~}

\documentclass[12pt,preprint]{aastex}

\slugcomment{To appear in ApJ}

\shorttitle{Small Scale Quasar Clustering}
\shortauthors{Infante et al.}

\begin{document}


\title{Quasars Clustering at $z\approx3$ on Scales $\lesssim 10 h^{-1}$ Mpc}


\author{L. Infante\altaffilmark{1}}
\affil{Departamento de Astronom\'{\i}a y Astrof\'{\i}sica, P. Universidad
Cat\'olica de Chile, Casilla 306, Santiago 22, Chile}
\email{linfante@astro.puc.cl}

\author{J. Varela}
\affil{Instituto de Matem\'aticas y F\'{\i}sica Fundamental, Consejo Superior
de Investigaciones Cient\'{\i}ficas, C/ Serrano 113 B, 28006 Madrid, Spain}

\author{M. Moles}
\affil{Instituto de Astrof\'{\i}sica de Andaluc\'{\i}a, Consejo Superior
de Investigaciones Cient\'{\i}ficas, C/ Camino Bajo de Hu\'etor 24, 18080
Granada, Spain}

\author{G. Hertling and A. Garc\'{\i}a}
\affil{Departamento de Astronom\'{\i}a y Astrof\'{\i}sica, P. Universidad
Cat\'olica de Chile, Casilla 306, Santiago 22, Chile}

\and
\author{F. Menanteau}
\affil{Department of Physics and Astronomy, Johns Hopkins University}


\altaffiltext{1}{Visiting Astronomer, European Southern Observatory.}


\begin{abstract}

We test the hypothesis whether high redshift QSOs would preferentially appear
in small groups or pairs, and if they are associated with massive, young
clusters. We carried out a photometric search for \Ly emitters on scales
$\lesssim 10 h^{-1}$ Mpc, in the fields of a sample of 47 $z\approx3$ known
QSOs. Wide and narrow band filter color-magnitude diagrams were generated for
each of the $6'.6\times6'.6$ fields. A total of 13 non resolved objects with a
significant color excess were detected as QSO candidates at a redshift similar
to that of the target. All the candidates are significantly fainter than the
reference QSOs, with only 2 of them within 2 magnitudes of the central object.
Follow-up spectroscopic observations have shown that 5, i.e., about 40\% of the
candidates, are QSOs at the same redshift of the target;  4 are QSOs at
different z (two of them probably being a lensed pair at z = 1.47);
2 candidates are unresolved HII galaxies at z$\sim$0.3; one
unclassified and one candidate turned out to be a CCD flaw. These data
indicate that at least 10\% of the QSOs at z$\sim$3 do have companions.

We have also detected a number of resolved, rather bright \Ly
Emitter Candidates. Most probably a large fraction of them might
be bright galaxies with [OII] emission, at z$\approx$ 0.3. The
fainter population of our candidates corresponds to the current
expectations. Thus, there are no strong indication for the
existence of an overdensity of \Ly galaxies brighter than m
$\approx$ 25 around QSOs at $z\approx$ 3.

\end{abstract}


\section{Introduction}

Quasars are among the most energetic objects known, and as such they can be
observed at very high redshift. It is therefore tempting to use them as
tracers of the properties of the distant Universe. However, their relation to
the physics of the far Universe is not yet well established, in particular their
link with galaxy formation and early evolution, and with density enhancements.
If there exist a measurable relation, then QSOs could be used as light houses
to study those phenomena at high redshift.

Whatever the mechanisms to form and feed QSOs are, it seems that
they would be associated with dynamical activity, merging in
particular (~\cite{kauffmann2000}), perhaps in regions of higher
than average density where the merging processes would be more
frequent. It is well known that at low redshift (z $\leq$ 0.6)
QSOs are not frequently found in pairs, and very seldom sitting in
clusters or dense groups (\citet{komberg94}; ~\citet{olmo91}).
However, the situation might be different at high-z, when massive
galaxies are forming and  merging processes are frequent. Indeed,
QSOs could be associated with massive primordial galaxies, or be
the result of violent interaction among galaxies (merging or
galactic cannibalism) which would be more frequent in the central
parts of massive clusters (\citet{komberg94}). If this is the
case, high redshift QSOs would tend to appear associated in pairs
or small groups, probably located in overdense regions.

There are a number of arguments which support those views.
Regarding the spatial QSO distribution, it is known that the QSO
2-point correlation length is similar to that of normal galaxies,
$r_o \approx$ 5-6h$^{-1}$ Mpc (~\cite{croom2001}), at all
redshifts (\cite{franca98}; ~\cite{croom2001}). Moreover, the
distribution of pairs of QSOs (with projected angular distance
$\theta < 10^o$) is similar to that of galaxy cluster pairs
(\cite{komberg94}). Results by different teams do support the idea
of some QSO grouping at high redshift. ~\cite{stephens97} found
three close quasar pairs at $z>2.7$ based on the Palomar Transit
Grism Survey. ~\cite{campos99}, have identified a number of \Ly
Emission Candidates (LECs) in a field that shows a decrement in
the CMBR temperature, supposedly produced by an unseen cluster at
z$\approx$2.5. Three of the identified objects in the field (of
which 2 were previously known) are QSOs at the same redshift. It
can also be mentioned the result by ~\cite{pentericci2000}, who
have found a number of LECs around a powerful radio galaxy. One of
the objects that has been confirmed spectroscopically is a QSO at
a redshift slightly higher than that of the radio-galaxy itself.
The last empirical argument is related to the density peak found
by ~\cite{steidel2000} at z$\approx$3. It also contains a QSO in
one of its sub-condensations.

It has been shown by ~\cite{tyson86} that the environments of
bright QSOs at  intermediate z values show an enhanced projected
galaxy density compared to the general field. Clustering at
redshift $\geq$2 is still poorly constrained by  observational
data. However, evidence for clustering is going to grow rapidly
with the advent of Large Telescopes. ~\cite{steidel2000} have
recently presented strong evidence of a significant concentration
of galaxies at z$\approx$3 based on Lyman-break  color selection.
~\cite{fynbo2001} and ~\cite{moller2001} have measured star
formation rates on seven \Ly emitting  galaxies located in a
filament at $z\approx3$, detected with narrow-band imaging
techniques, and confirmed spectroscopically.

Direct searches have not yet given definitive answers
(~\cite{keel99}, and  references therein). The identification of
luminous objects associated with  forming clusters would be of
great help in the study of matter at high redshift.  If the nature
of QSOs or, in general, very energetic objects, is somehow related
to the density of their environment (at least in the early epochs
of galaxy formation), then would we expect small scale clustering
of QSOs, and  clusters around every QSO? A question that is
difficult to unambiguously answer by means of previous QSOs
catalogs, as they are based on surveys that search  for the
presence of radio emission, UV excess or strong emission lines.
Here we address this question directly.

In this work we present the results of a photometric search for
LECs around 47 QSOs at z$\sim$3. We also report the spectroscopic
follow up of 12 of the 13 unresolved candidates we found. The
observational strategy used to look for candidates is discussed in
Section~2. The results are presented and discussed in Section~3.
In order to consider QSOs separately, we pay special attention in
classifying stellar and non-stellar candidates. The discussion and
summary are given in Section~4.

\section{Observations and data reduction}

In order to study the clustering properties of high  redshift QSOs
at small  scales, and their possible association with galaxy
clusters, we carried out a  systematic search for strong \Ly
emitters in the neighborhood of known QSOs.  Based on the VLT
filter availability (see Table~\ref{filters}), we selected all
known QSOs in the southern hemisphere, with redshift within the
ranges  2.863$\leq$ z $\leq$ 2.959, and 3.090$\leq$ z $\leq$
3.222. Our initial target  list included 27 and 36 objects
respectively, with RA between 21h and 5h.

\subsection{Observational Strategy}

The combination of a narrow band filter (\n) at the wavelength of
redshifted \Ly, and a wide filter (\w) to measure the continuum
has been used by numerous authors to detect LECs (see
~\cite{steidel2000}, and references therein). By defining the zero
point of the \wn color index as that corresponding to pure
continuum objects, any significant, positive color excess serves
to identify LECs.  That zero point is dependent on both the
filters and the colors of the pure continuum objects.

To calculate the optimal filter combinations, take $f(\lambda)$ as
the flux distribution of a given object, and $S_w(\lambda)$ and
$S_n(\lambda)$ the filter transmission functions in the wide and
narrow filters respectively. The \wn color index is given by

\begin{center}
\( w-n=-2.5log\left( \frac{\int f(\lambda )S_{w}(\lambda )d\lambda
}{\int S_{w}(\lambda )d\lambda }\right) +2.5log\left( \frac{\int
f(\lambda )S_{n}(\lambda )d\lambda }{\int S_{n}(\lambda )d\lambda
}\right)  \)
\end{center}

For a pure, flat continuum object, \wn is obviously zero. For some fixed,
non-flat continuum, the value of \wn will be at its minimum when both filters
are centered at the same wavelength. Therefore, to minimize color effects, we
choose filter sets such that the narrow one is always contained in the broad
one (see table~\ref{filters}).

With the above definition of \wn, an object with emission lines in the
redshift range surveyed by the narrow filter (approximately given by
its FWHM) would  have a positive color index. And all the objects with
emission lines within the range of the broad filter, but out of the
narrow one, would produce a negative  color index. In fact, the
presence of absorption lines would produce the  opposite
effect. However, to produce a significantly positive color index, that
absorption in the wide filter should be very strong to compensate for
the width of the filter.

\begin{figure}
\epsscale{0.7} \plotone{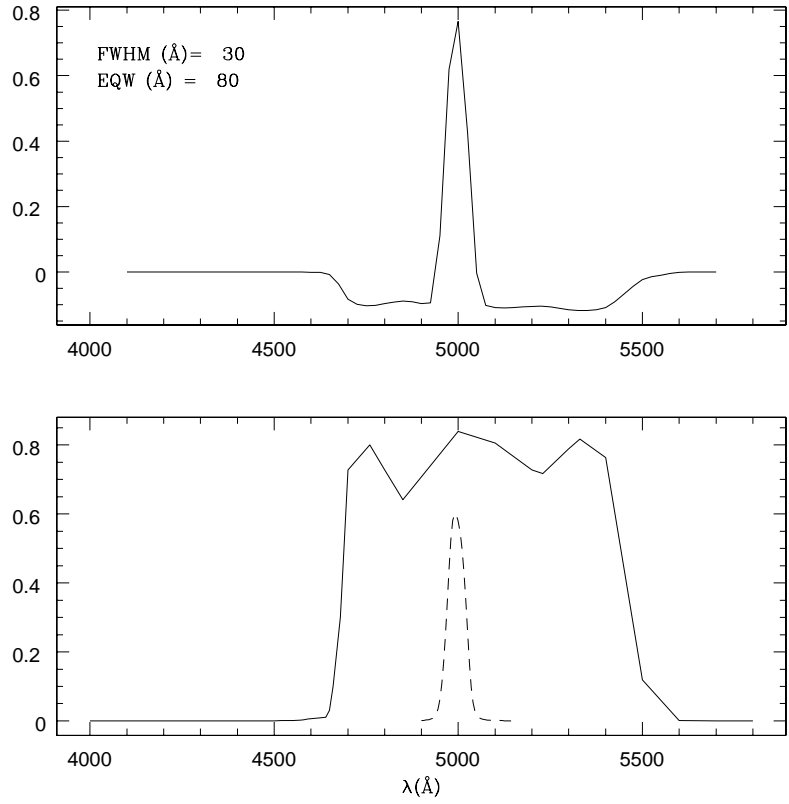} \caption{The expected signal for
an object with \Ly in emission, as a function of the redshift. The
curves correspond to the $g_{Gunn}$ and $[OIII]500.1$ filter
combination. \label{simul_30_80}}
\end{figure}

To illustrate the expected \wn colors, we present in
Figure~\ref{simul_30_80}  the results of color simulations carried
out for one of the filter combinations.  Notice that other than
the zero-color sequence produced by the pure continuum  objects in
the field, there is a second quasi-sequence, with a negative color
index. This is due to objects with emission lines out of the range
of the narrow  filter, but within the broad one. The shape of the
sequence depends on the shape of the broad band filter. If the
filter is a top-hat shaped filter the effect  of the emission line
is almost independent of its position within the filter,  except
when near to the edges. Simulations also show how color indices
are distributed. It is clear that the precise value of the color
index will depend, other than on the redshift and the strength of
the line, on its FWHM. When the emission line is broader than the
narrow filter, the color index is smaller.

\subsection{Contamination and losses}

\begin{figure}
\epsscale{0.9}
\plotone{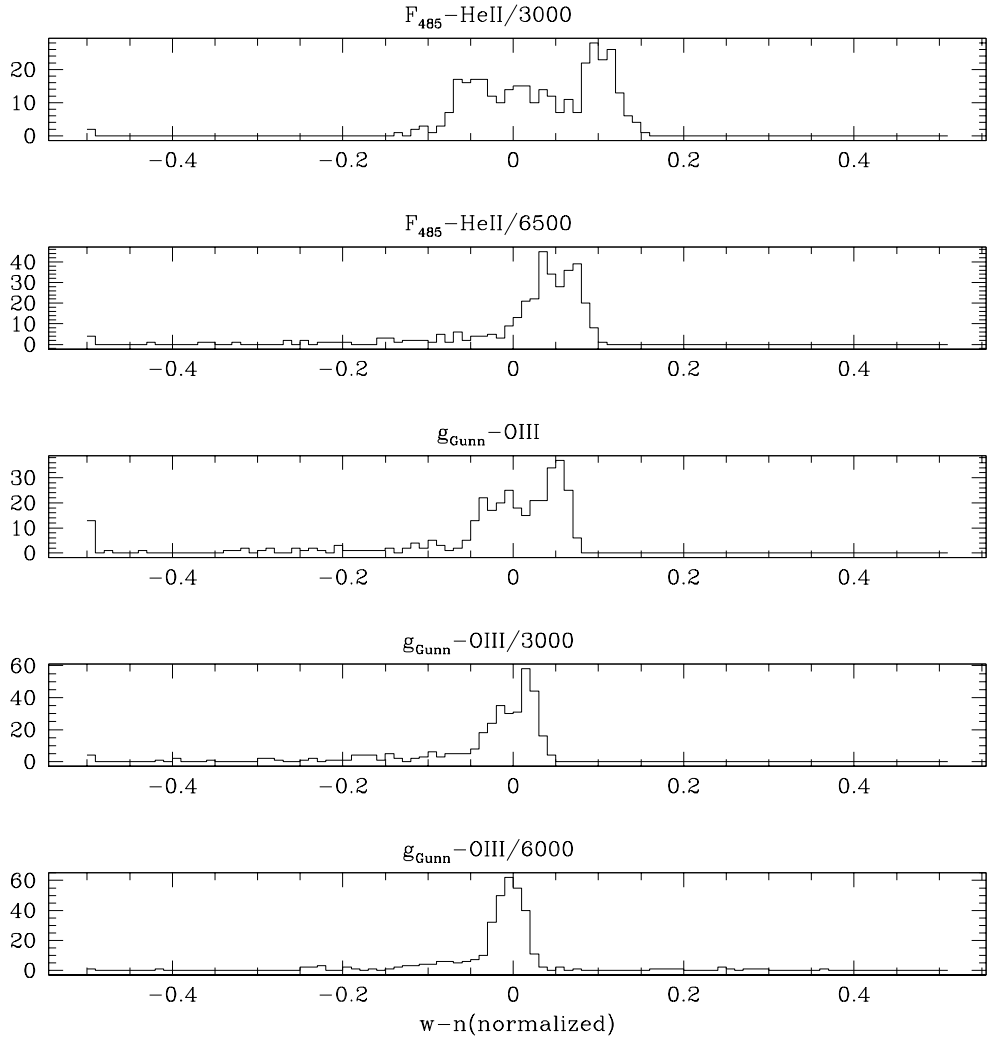}
\caption{Distribution of w-n for a sample of stars for each set of filters
\label{starcolor}}
\end{figure}

When looking for QSO-candidates, one potential problem is the
misidentification of stars as QSOs. Depending on the filters and
the spectral type, a star can  produce a positive \wn color index
that could be misclassified as a  $z\approx3$ quasar candidate. In
order to estimate the number of misidentifications we have
calculated the \wn index for different types of stars using two
spectral libraries (\cite{gunn83}; ~\cite{jacoby84}). These
libraries include spectral classes from O to M, and luminosity
classes from III to V. The results are illustrated in Figure
~\ref{starcolor}, which shows that the  stellar colors are well
grouped around zero, and that a strict upper limit can be defined.

Another consideration is that some real QSOs may be lost if they
don't produce a  high enough color index. Indeed, a \Ly emission
QSO can produce a zero or even  negative color index depending
upon its characteristics and the precise position of the line in
the spectral region covered by the filters. In fact, some of our
target QSOs do have such color indices. Examination of their
spectra shows that in most cases this is due to the presence of a
strong absorption in the region of the \Ly line. It appears that
these types of losses can be as high as $50\%$ when the filter
combination $F_{485}-HeII/3000$ is used, whereas it is only $20\%$
for the $g_{Gunn}-[OIII]$ combination. For the other filter
combinations the losses are negligible.

Although not a serious problem, other emission line objects or
QSOs at other redshifts can be confusing when  lines other than
\Ly fall in the region covered by  the narrow band filter. For
unresolved candidates, a first approximation to the contamination
effect can be estimated using the 2QZ catalog
\footnote{\footnotesize The 2dF QSO Redshift Survey (2QZ) was
compiled by the 2QZ survey team from observations made with the
2-degree Field on the Anglo-Australian Telescope.}. We give in
Table \ref{lowerz} the redshift range  and the lines that could
produce color indices that mimic those of a z = 3 QSO. From the
same 2QZ catalog we find that the number of QSOs in the other
redshift ranges amounts to about 4 per square degree, down to the
catalogue limit  of B$\sim$21. Therefore, contamination by bright
QSOs should not be a problem, with only 1 or 2 expected in our
field of view. Given the magnitude limit of our survey, and
assuming an homogeneous distribution of QSOs, a total of at about
8 QSOs at redshifts different from the central one would be
expected in our case.  As discussed before, some candidates can go
undetected depending on different  aspects and, therefore, those
numbers should be considered as upper limits. All in all, it is
unlikely that our list of point-like candidates could be heavily
contaminated by stars or by bright QSOs at a different redshift,
even if the contamination by fainter QSOs could be sizable.

It is also possible that some unresolved HII galaxies at
z$\approx$0.3 could  be falsely identified as QSO candidates. This
effect cannot be properly estimated since very little is known
about the distribution  of compact HII galaxies. The situation is
different for resolved candidates, since the effect of the [OII]
line of faint HII galaxies at z$\approx$0.3 could be mistaken as
due to LECs. We will discuss further this aspect below.

\subsection{Imaging}

Observations were performed with FORS1 at ANTU VLT on the nights
of 22 and 23  October, 2000. Observing conditions were clear and
photometric with seeing  between 0\farcs4 and 1\farcs0. The CCD
camera at FORS1 has a pixel size of  0\farcs2/pix, giving a total
field of view of 6\farcm6$\times$6\farcm6. For  each QSO we
acquired images with a narrow band filter covering the wavelength
of the redshifted \Ly line, as well as a broader band filter to
get the \wn  color index. We exposed $3\times20$ seconds with the
broad band filters ({\sl  special 485} ($F_{485}$) and g-Gunn
respectively) and $3\times200$ seconds with the narrow band
filters (HeII~472.6, HeII~ 478.1, [OIII]~500.1, [OIII]~504.5,
[OIII]~510.5). Landolt standard stars were observed at various air
masses to  calibrate the data. Extinction terms and their
uncertainties are presented in Table~\ref{filters}. In 2 nights at
the VLT we observed 47 fields, i.e. a number of objects large
enough to allow a meaningful statistical analysis.

Image detection, photometry and classification were performed
using the Source  Extractor (SExtractor) software program
\citep{bertin96}. Image analysis  proceeded as follows: CCD images
were sectioned in sub-areas of $64\times64$ pixels. A background
map was constructed. The local background histogram was clipped
iteratively until convergence at $3\sigma$ was reached. These raw
background values were then filtered using a median filter (in
3$\times$3  sub-areas), and a bi-cubic-spline was used to obtain
the resulting background  map. The background map was then
subtracted from the image. All objects with a  threshold $\ge$
1.5$\sigma$ above the sky level, and with a minimum area of 5
pix$^{2}$ were selected and extracted. Their total magnitudes were
calculated by integrating the intensity within elliptical
apertures. Finally, the star/galaxy classification
(non-resolved/resolved objects) was done with a neural network
routine developed by \citet{bertin96}. This is a training program
that  classifies objects with an index (stellarity index) between
0 (galaxies), and 1 (stars). It uses eight isophotal areas, the
peak intensity, and the seeing  information to classify the
objects. All the objects that SExtractor had no  problem with
their photometry (i.e. SExtractor's \tt FLAG \rm $= 0$) were
selected.

The zero point of the broad band magnitudes was set to 27, which
results in QSOs magnitudes close to their catalog V magnitudes.
The color index zero point was set after examining the resulting
color-magnitude diagrams, as corresponding to the sequence seen in
the simulations. The threshold to accept the color index of a non
resolved candidate as positive was established after considering
two effects. First, we took into account the width of the band
defined by the stellar colors. They are different for the
different filter combinations: 0.15 mag for the F485-HeII/3000
set, 0.10 for F485-HeII/6500, Gunn~g-[OIII]/50 and
Gunn~g-[OIII]/3000 sets; and 0.05 for the Gunn~g-[OIII]/6000 set.
Second are the observational errors. There are well defined trends
of errors with the broad-band magnitude, with 1$\sigma$ errors of
0.18~mag at \w = 23.5. We decide however to use the error of each
individual object. The reason is that, for a given flux in the
broad filter, the LECs will have in the narrow filter a higher
signal than the non emitting objects. Thus, adopting instead the
average error trend would penalize the detection of LECs.
Unresolved objects were finally flagged as LECs when their color
index exceeded the sum of the corresponding width of the stellar
sequence, and 3.5 times the error of their \wn index.

Resolved objects were considered as candidates when their color
indices exceeded than 5 times the error in their measured color.

\subsection{Spectroscopy}

Long slit observations of 12 QSO candidates were carried out on 3
nights in  September 2002, using LDSS2 on the Baade telescope at
the Las Campanas Observatory. The slit width was 1\arcsec and the
CCD pixel size was $0.380~\mu m$. The  slit was aligned along the
line connecting either two member candidates or a  candidate and
the central QSO. Exposure times ranged from 1800 to 3600 seconds.
The data were taken in non photometric weather and the seeing
varied from 0\farcs6 to 1\farcs4.

The spectra were reduced  with IRAF. The data were first bias
subtracted and flat-fielded. Sky removal was done by subtraction
of  sections adjacent to the  spectrum in the slit. Wavelength
solutions were calculated from Helium-Argon arc lamp exposures.
The rms fit of the arc lines given by IRAF is $ 0.5 {\AA \,
pixel^{-1}}$ while using a $4^{th}$ order polynomial to fit
approximately  25 lines. The wavelength coverage is approximately
$3700 {\rm \AA}$ to $7500 {\AA}$ with a linear reciprocal
dispersion of $5.3 {\AA \,/pixel}$. Cosmic rays were removed via
the COSMICRAY routine.

\section{Results}

\subsection{Imaging}

\begin{figure}
\epsscale{1} \plotone{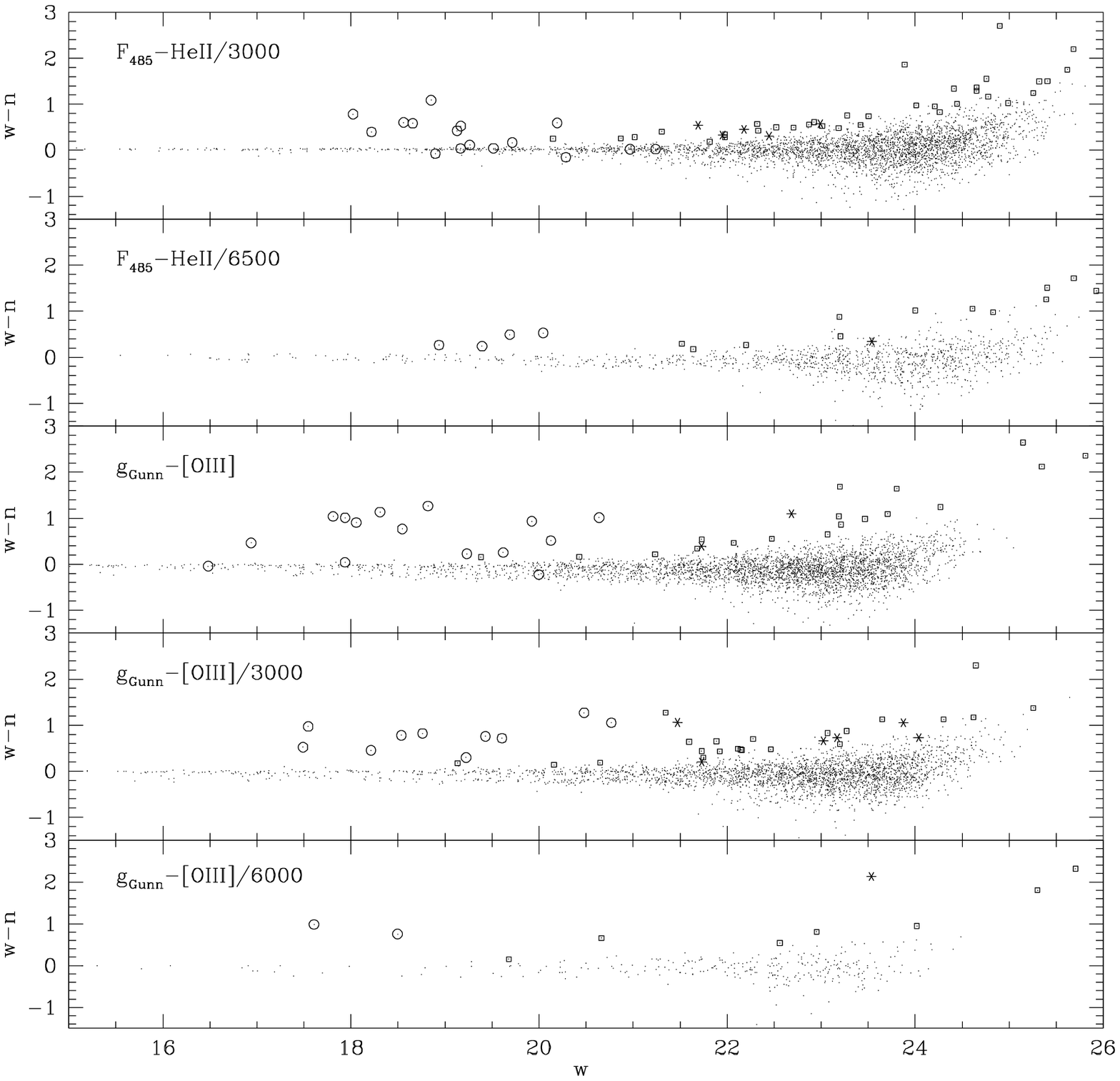} \caption{Color - magnitude diagram
for each set of filters. Circles are central QSOs, stars are
unresolved candidates, boxes are resolved candidates, and points
are all detected sources. \label{colormg1}}
\end{figure}

\begin{figure}
\epsscale{1} \plotone{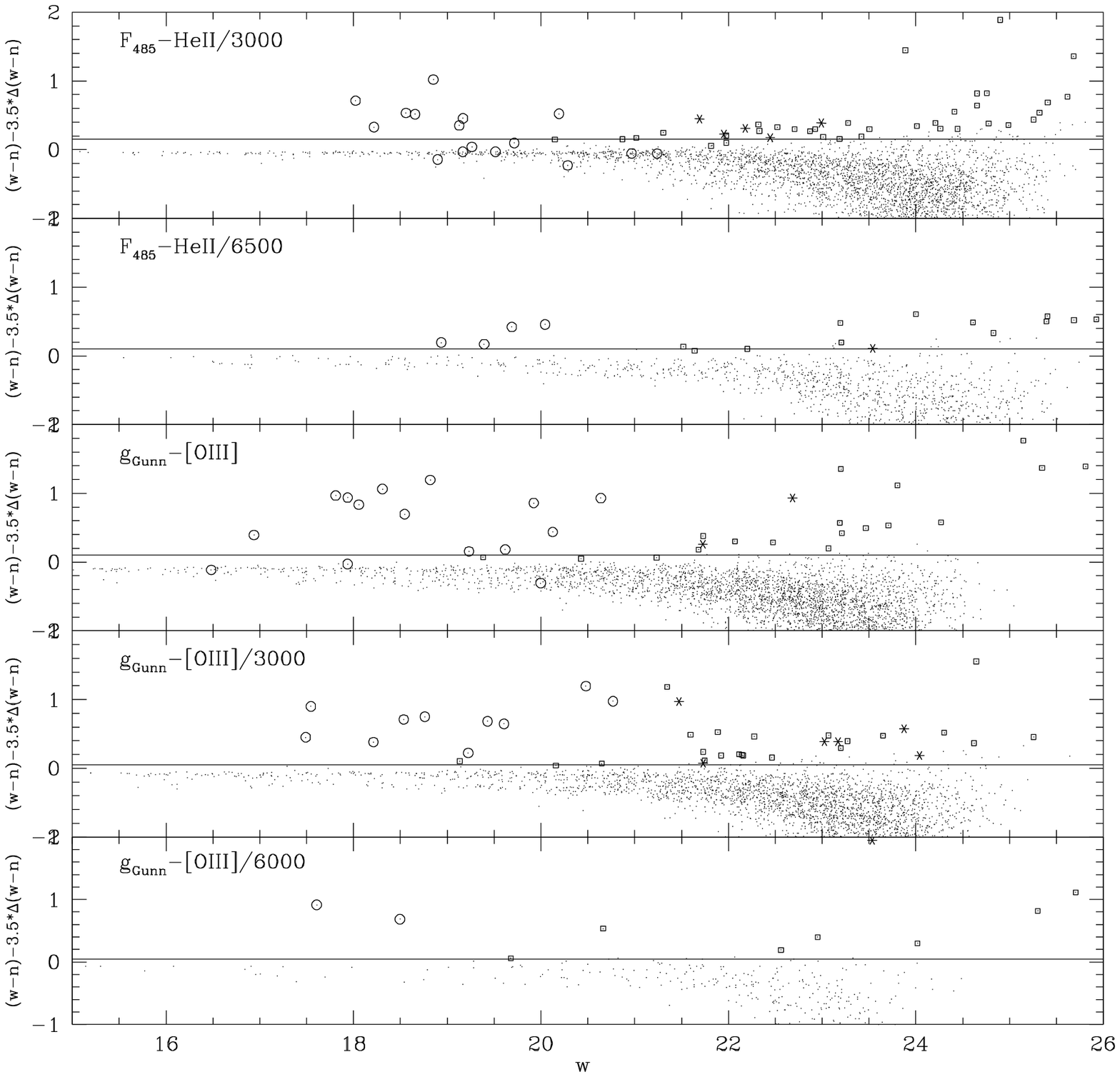} \caption{Color detection threshold -
magnitude diagram for each set of filters. Circles are central
QSOs, stars are unresolved candidates, boxes are resolved
candidates, and points are all detected sources. \label{colormg2}}
\end{figure}

Figure ~\ref{colormg1} shows the color index \wn {\sl versus} the
broad-band magnitude, \w, and Figure ~\ref{colormg2} gives the
detection threshold, (\wn-3.5$\sigma$), {\sl versus} \w. We have
plotted together all the fields observed with the same filter
combination. As discussed in the previous sections, \wn is
positive for objects with some emission line lying within the
narrow band filter, and negative when an emission line falls in
the broad band filter. From the previous considerations, all
objects with \wn over the defined threshold are LECs in the
redshift range determined by the width of the narrow band filter.
The unresolved LECs are QSO candidates at the same redshift of the
central QSO. Resolved objects could be galaxies at the central QSO
redshift. The images of all the LECs were visually inspected on
the processed images. If the object presented distortions, or if
it was too close to the edge of the frame, or if a nearby object
might be affecting the photometry, then it was eliminated from the
list of candidates.

Adopting a flat cosmological model with H$_0$ = 70~\ksM,
$\Omega_{m}$ = 0.3, $\Omega_{\Lambda}$ = 0.7, the size of the
field covered by our frames corresponds to 12.2$\times$12.2
Mpc$^2$, comoving size, \ie, about 1.5 times the clustering scale
length of QSOs found by \cite{croom2001}. In the radial direction,
the redshift range covered by each of our narrow band filters
translates into $\sim$50 Mpc.

\subsubsection{QSO candidates}

The main results concerning unresolved LECs, are presented in
Tables ~\ref{companions}, ~\ref{stat_qsos}, and
~\ref{statfield_qsos}. We found a total of 13 QSO-like LECs in 11
fields. In two fields (\#14 and \#43) there are two candidates. As
we already argued, this is a lower limit to the number of
candidates. It is interesting to note that those companions have
broad-band magnitudes significantly fainter than the central QSO.
Therefore, if there is some grouping of QSOs at high redshift, the
members have significantly different luminosities.

\begin{figure}
\epsscale{1} \plotone{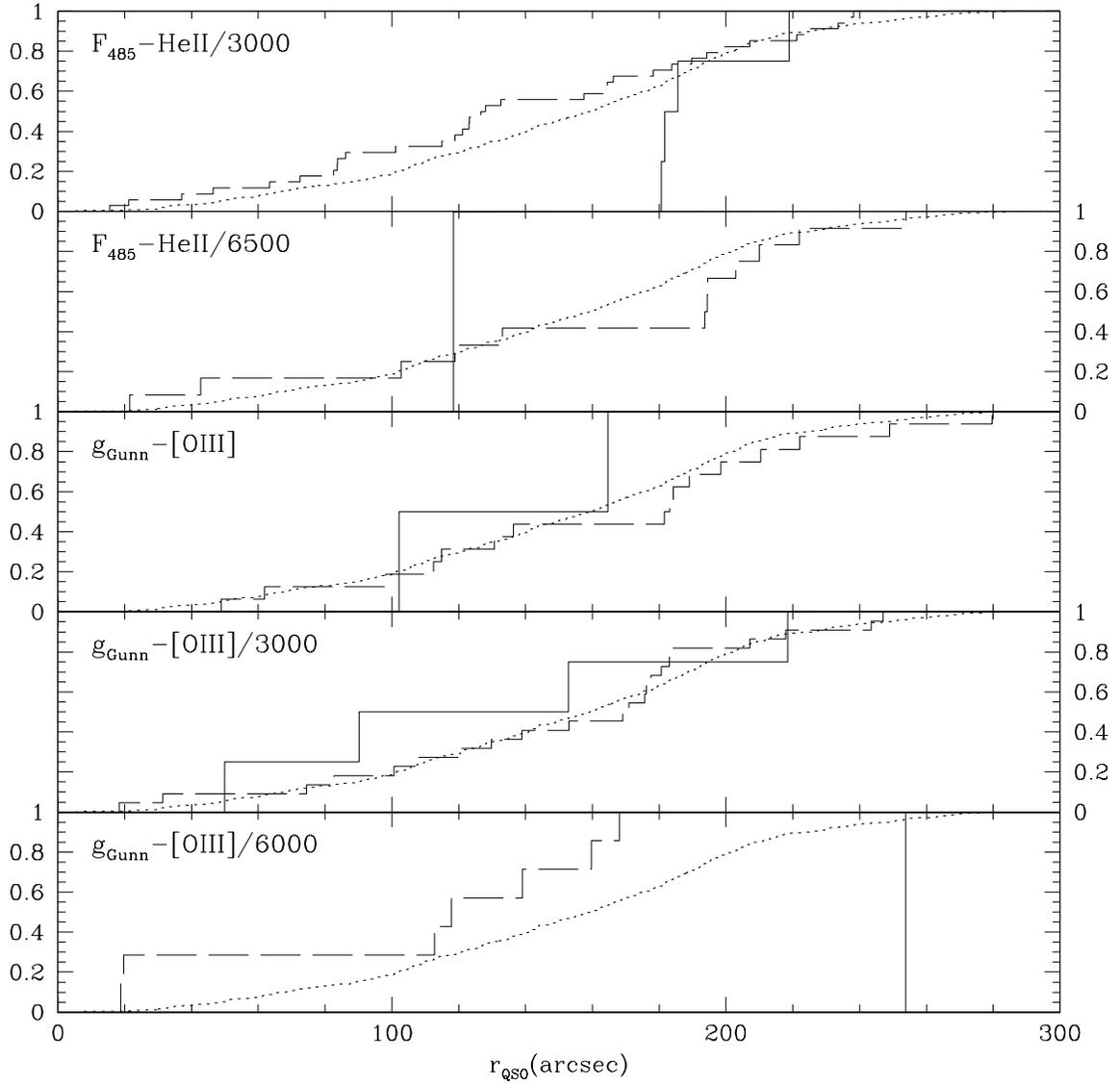} \caption{Radial distribution of LECs
for each filter combination. Solid lines are for unresolved
objects, and dashed lines are for resolved objects. The dotted
lines illustrate what would be expected for randomly distributed
objects. \label{radialdist}}
\end{figure}

We have also considered the spatial distribution of the
QSO-candidates around  the central QSO. In Figure~\ref{radialdist}
we present the cumulative,  normalized distribution of the radial
distances from the QSO-candidates to the  targets for each set of
filters. Given the small number of candidates no firm  conclusion
can be extracted; however, the candidates appear to be closer than
would be expected for a random distribution.

\begin{figure}
\epsscale{1} \plotone{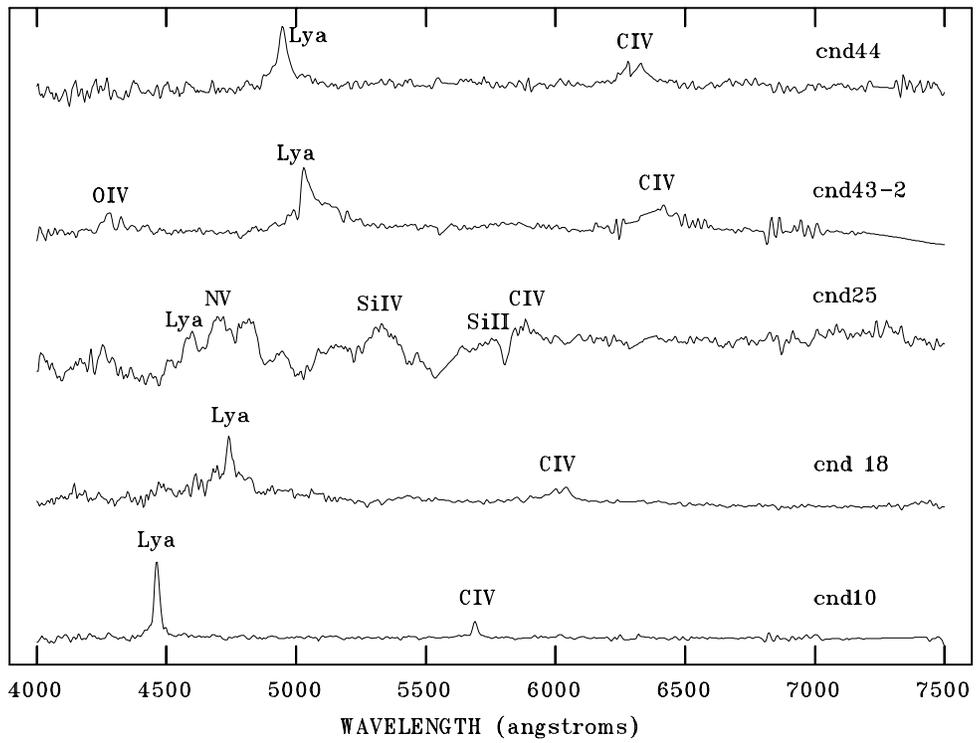} \caption{Confirmed $z\approx3$ QSO
candidates at the host QSO redshift. The main QSO emission lines
are identified. \label{spec1}}
\end{figure}

\begin{figure}
\epsscale{1} \plotone{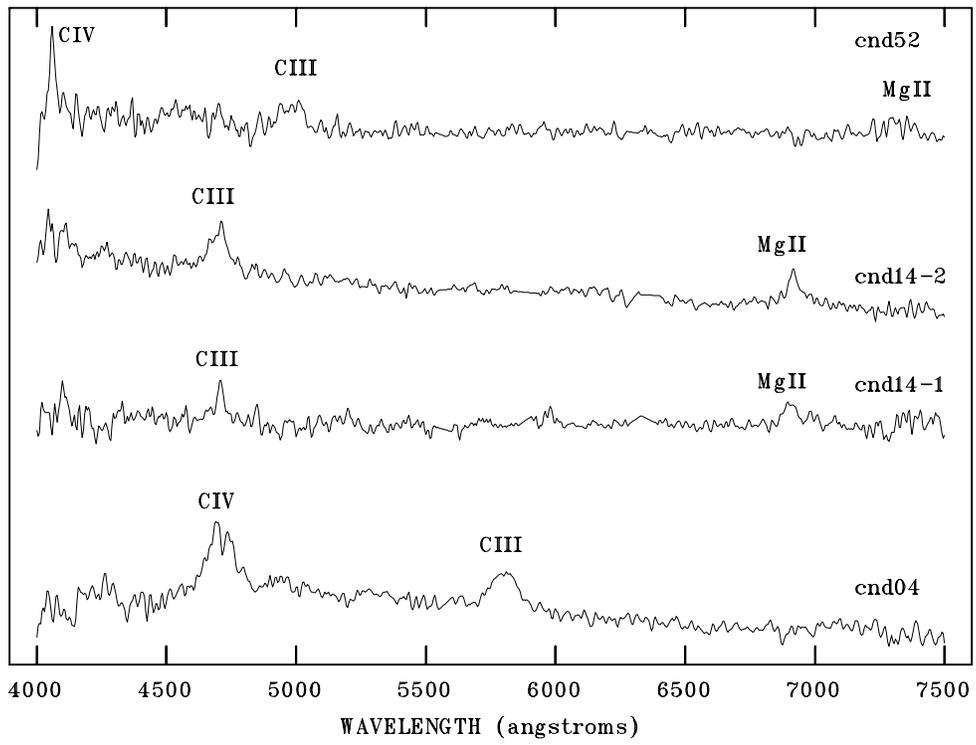} \caption{QSOs or AGNs not at the
host QSO redshift. The main emission lines are identified.
\label{spec2}}
\end{figure}

\begin{figure}
\epsscale{1} \plotone{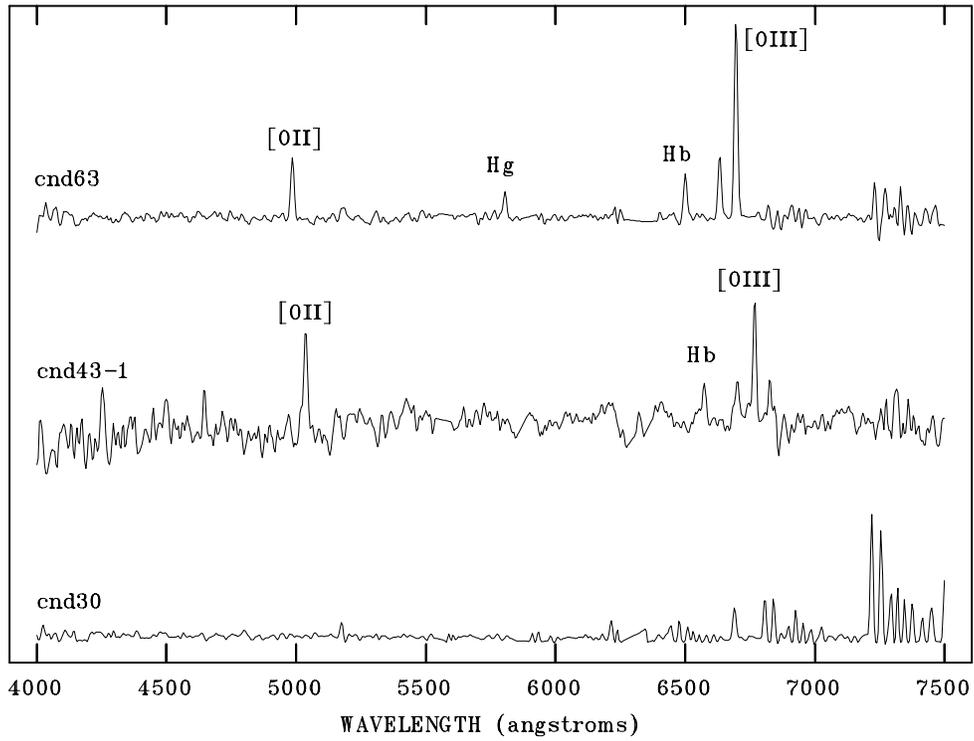} \caption{Non QSO objects. Most
probably, the two top spectra correspond to starburst galaxies.
The bottom  spectrum has not been identified. The main emission
lines are identified. \label{spec3}}
\end{figure}

Spectra for 12 of the 13 QSO candidates and 5 central QSOs were
obtained, although one candidate turned out to be a CCD cosmetic
defect. After extraction, wavelength  reduction and flux
calibration, we proceeded to classify the spectra and measure
redshifts. The resulting spectra are presented in
Figures~\ref{spec1},~\ref{spec2} and \ref{spec3} and their
classification are listed in Table~\ref{spec-class}.

As a result, 5 out of 12 candidates show Ly$\alpha$ in the narrow band filter,
therefore, these are QSOs at the central QSO redshift. Thus, as argued before,
a minimum of 10\% of the surveyed z$\sim$3 QSOs have companions at the same
redshift and, therefore, pairing of QSOs at high redshift would be not
infrequent.

Four of the other candidates are also QSOs, but at  different
redshifts. Two in the field of the QSO \#14 have very similar
spectra and could be a lensed pair. Two unresolved candidates
turned out to be faint HII galaxies at z$\sim$0.3. The last
candidate spectrum has too low signal to noise ratio for a proper
classification.

\subsubsection{Resolved candidates}

Although our main interest was the clustering of QSOs, we have
also found a number of resolved LECs. Tables \ref{stat_res} and
\ref{statfieldres} list the main results determined from the 96
resolved LECs  in 34 fields. These data shows that the number of
candidates brighter than m = 23.5 is much higher than expected
from the measured Luminosity Function (LF) of \Ly Break Galaxies
(LBG) at z$\approx$3  (~\cite{Adelberger2000}). This excess become
even more significant when one takes into account that the number
of  LECs found represent only a small fraction  (20\%) of the LBGs
that would also be \Ly emitters.

The most plausible explanation is that most, or even all, of those
objects are galaxies at z$\approx$ 0.3, with the [OII] line
falling in the narrow band window. Many of these objects are in
fact just barely resolved, probably corresponding to HII galaxies
which have a characteristic faint continuum with prominent
emission lines (see ~\cite{T91}). The [OII] line is narrow and
intense in those objects. Essentially all the objects with an
[OII] EW over 60~\AA, the median value for the galaxies in
Terlevich et al. (1991), would be detected as candidates.

Moreover, four of the bright LECs are found grouped in a field, at
the position of a known radio galaxy (NVSS-J104545-012252), with
which they could define a cluster at redshift $\approx$ 0.3.

In the range 23.5 $\leq$ m $\leq$ 25, the expected number of \Ly
emitters is close to what we have found. Therefore it doesn't seem
that there is a significant overdensity of \Ly emitters around
QSOs, in that magnitude range. Fynbo et al (2001) have suggested
that there may be an excess of fainter  \Ly emitters around some
QSOs, however, we cannot address this issue given the magnitude
limit of our survey.

Figure ~\ref{radialdist} shows the radial distribution of all the
resolved candidates. This distribution is not far from what would
be expected for a random distribution. When only the faint
candidates are taken into account, the result is even closer to
the expectation for a random distribution.

\section{Discussion and Summary}

We have explored the use of narrow band filters together with
broad band filters to detect \Ly emitters around 47 known QSOs.

For unresolved candidates we studied possible contamination from
stars by simulating \wn colors from stellar spectral libraries
covering a wide range of spectral types and luminosities.
Contamination by objects with emission lines at redshifts $\neq3$
was also analyzed.

As a result of the survey we found a total of 13 unresolved LECS
in 11 fields. The possible companions to the targeted z$\approx$3
QSOs are systematically fainter, with only 2 objects within 2
magnitudes of the central QSO. The selection criterion make it
very unlikely the these objects are stellar contaminants.
Moreover, from the 2QZ catalog we find that contamination by
brighter than B = 21 mag QSOs at a different z, but with strong
lines entering the narrow filter, is very small; none of our
candidates are brighter than this limit.

In principle, our candidates could  be unresolved Ly-break
galaxies (LBG) instead of true QSOs. Before looking at the
spectroscopic data, some arguments can be given:

(i) Given the {\em angular size - redshift}t relation, for the
geometry we use  in this paper, the scale at $z=3$ is 7.7
kpc/\arcsec. Thus 0\farcs4 -  0\farcs8 corresponds to linear sizes
in the range between 3 and 6 kpc.  At these scales, only very
small or extremely nucleated galaxies would appear as unresolved.

(ii)At $z\approx3$ and $V<24$, the number of galaxies is very
small. As a rough  estimate we use the ~\cite{Adelberger2000}
luminosity function to calculate the  number of Ly-break galaxies
at z = 3. Assuming that 20\% are emission line  galaxies, given
our geometry and a
$50~Mpc\times12.2~Mpc\times12.2~Mpc\times47~fields\simeq3.5\times10^5
~Mpc^3$ volume, we expect 0.0002 ($R<22.5$), 1.4 ($R<23.5$), 5.1
($R<24$) and 18.5 ($R<24.5$) Ly-break emission galaxies. However,
we find 7, 11, 12 and 13 unresolved objects respectively. Our
photometric bandpass is different, but this should not
significantly affect these calculations. Therefore our results
show a clear excess of unresolved bright objects.

The spectroscopic data confirm that 9 of our 12 candidates are
QSOs, 5 of  them at the redshift of the target QSO, at z$\sim$ 3.
From the 2QZ catalog we  find that only a very tiny fraction of
QSOs have a companion in a volume similar to that of our survey.
Due to the lack of data it is hard to extend the analysis to
fainter magnitude  limits. Nevertheless, starting form the
luminosity function of faint AGNs given by Pei (1995, see their
Fig. 8), it is possible to deduce that in the range $2.2 < z <
3.3$, there are $\sim 30$ objects  $deg^{-2}~mag^{-1}$ at
$B\sim22.5$, and $\sim 10$ objects $deg^{-2}~mag^{-1}$ at
$B\sim21.0$. The redshift range covered in our survey is 0.05,
which is $\sim$ 1/22 of Pei's range. For the area covered by our
survey, the expectations would then be
$N_{expected}(B\sim21)\approx 0.22$ and $N_{expected}(B\sim22.5)
\approx 0.72$, and we find 0 and 3 respectively. By extrapolating
Pei's curve and integrating between 22 and 24, we would expect
2.75 QSOs, and we have found 5. Taking into  account the preceding
numbers and the possibility to have lost other candidates, as
previously discussed, we can conclude that our findings indicate
an excess of  QSOs around QSOs at z$\sim$3.

The spectroscopic results  confirm the reliability of our
photometric technique to detect QSO companions. Nearly half of the
candidates have been confirmed as QSOs at the same redshift of the
central object, and none were found to be stars.

Our search has also produced a number of resolved LECs. We argue
that the  majority with m$\leq$23.5 are probably galaxies with
[OII] emission at z$\approx$ 0.3, many of them being HII galaxies.
In particular, 4 of the resolved LECs appear grouped in one field,
close to a known radio galaxy (NVSS-J104545-012252), and  could be
part of a larger group at that redshift.

The density of resolved LECs in the range 23.5 $\leq$ m $\leq$ 25
is close to the expectations from the LF given by
~\cite{Adelberger2000} for LBG, provided that about 20\% of them
are also \Ly emitters, as pointed out by ~\cite{steidel2000}. The
radial distribution of detected LECs around the target QSOs is
very close what would be expected for a random distribution.

\newpage
\acknowledgments{

We acknowledge the {\em Proyecto de Cooperaci\'on
CONICYT(Chile)/CSIC(Espa\~na)} for partial financial support. LI
thanks {\em proyecto FONDAP and P. Universidad Cat\'olica de
Chile} for funding part of this project. J. Varela acknowledges a
grant FP98-52349748 from the {\em Ministerio de Ciencia y
Tecnolog\'{\i}a } of Spain. This research was partially funded by
the Spanish {\em Ministerio de Ciencia y Tecnolog\'{\i}a}, grant
PB98-0139.

This research benefitted from the use of the VizieR catalog access tool and
Aladin, CDS, Strasbourg, France. This research has made use of the NASA/IPAC
Extragalactic Database (NED) which is operated by the Jet Propulsion Laboratory,
California Institute of Technology, under contract with the National
Aeronautics and Space Administration.

We would also like to thank P. Hall and E. P\'erez for helping us
with the quasar identification, and Andrew Stephens and Eric
Gawiser for a careful reading of the manuscript.}

\clearpage


\begin{deluxetable}{l|ccc}
\tablecaption{Filters and Extinction Coefficients. \label{filters}}
\tablewidth{0pt}
\tablehead{
\colhead{Filter} &
\colhead{$\lambda_0~[\AA]$}   &
\colhead{FWHM~$[\AA]$} &
\colhead{$k$} \\
}
\startdata
$g_{Gunn}$     & 5060  & 795  & 0.137 $\pm 0.016$ \\ $F_{485}$      &
4850  & 370  & 0.176 $\pm 0.014$ \\ $HeII/3000$    & 4726  &  58  &
0.219 $\pm 0.028$ \\ $HeII/6500$    & 4781  &  68  & 0.211 $\pm 0.022$
\\ $[OIII]$       & 5001  &  57  & 0.187 $\pm 0.030$ \\ $[OIII]/3000$
& 5045  &  59  & 0.186 $\pm 0.036$ \\ $[OIII]/6000$  & 5105  &  61  &
0.164 $\pm 0.020$ \\
\enddata

\end{deluxetable}

\clearpage


\begin{deluxetable}{ccc}
\tablecaption{Lines that could fall into the narrow band filters
for redshifts lower than 3. \label{lowerz}} \tablewidth{0pt}
\tablehead{ \colhead{Line} &
\colhead{Ranges of redshift} \\
}
\startdata
$[OIII] \lambda4959\AA $& $ 0 - 0.042 $\\
$H\beta \lambda4861\AA $& $ 0.017 - 0.062 $\\
$H\gamma \lambda4340\AA $& $0.076 - 0.117 ; 0.139 - 0.190 $\\
$[OII] \lambda3728\AA $& $0.252 - 0.301 ; 0.326 - 0.386$\\
$MgII \lambda2803\AA $& $0.665 - 0.730 ; 0.764 - 0.843 $\\
$CIII] \lambda1909\AA $&$ 1.445 - 1.540 ; 1.590 - 1.706 $\\
$CIV \lambda1549\AA $&$ 2.013 - 2.130 ; 2.192 -2.335 $\\

\enddata

\end{deluxetable}

\clearpage


\begin{deluxetable}{cccccc}
\tabletypesize{\scriptsize} \tablecaption{Unresolved companions
\label{companions}} \tablewidth{0pt} \tablehead{ \colhead{$Filter
Set$} & \colhead{$\# field$} & \colhead{$w$} &
\colhead{$w-w_{QSO}$} & \colhead{$w-n$} &
\colhead{$r_{QSO}~(arcsec)$}
\\
}
\startdata

$F_{485}-HeII/3000$     & 04  &  21.95 & 0.97  &  0.33   & 219.0 \\
                        & 14  &  22.45 & 3.89  &  0.31   & 180.7\\
                        & 14  &  22.99 & 4.43  &  0.57   & 182.5\\
                        & 18  &  21.69 & 1.97  &  0.55   & 185.6\\
                        & 25  &  22.18 & 3.05  &  0.45   & 181.7\\
\hline
$F_{485}-HeII/6500$     & 10  &  23.54  &  3.50 & 0.98   & 118.5 \\
\hline
$g_{Gunn}-[OIII] $      & 52  &  21.73 & 4.79  &  0.39   & 164.6 \\
                        & 63  &  22.69 & 3.07  &  1.09   & 85.7\\
\hline
$g_{Gunn}-[OIII]/3000$  & 30  &  24.04 & 3.56  &  0.72   &  90.2 \\
                & 43  &  21.47 & 3.92  &  1.06   & 152.9 \\
                        & 43  &  23.17 & 5.62  &  0.73   & 50.0 \\
                        & 44  &  21.73 & 2.97  &  0.20   & 218.4 \\
\hline
$g_{Gunn}-[OIII]/6000$  & 49 &  23.53 & 5.03  &  0.88   & 253.8 \\
\enddata

\end{deluxetable}

\clearpage


\begin{deluxetable}{cccccccc}
\tabletypesize{\scriptsize} \tablecaption{Statistics by number of
unresolved companions \label{stat_qsos}} \tablewidth{0pt}
\tablehead{ \colhead{Filter Set} & \colhead{$N(w<21.5)$} &
\colhead{$N(w<22.5)$} & \colhead{$N(w<23.5)$} &
\colhead{$N(w<24.5)$} & \colhead{$N(Total)$} & \colhead{\(
\frac{N}{N_{fields}} \)} & \colhead{\( \frac{N}{d\Omega/dz} ~
(\#/\square') \)}
\\
}

\startdata

   $    F_{485}-HeII/3000  $           &       0     &       4     &       5     &       5     &       5   &    0.31   &   0.15 \\
   $                       $          &      0.0 \%     &     80.0 \%     &    100.0 \%     &    100.0 \%     &    100.0 \%   &  &   \\

   $    F_{485}-HeII/6500  $           &       0     &       0     &       0     &       1     &       1   &    0.25   &   0.10 \\
   $                       $          &      0.0 \%     &      0.0 \%     &      0.0 \%     &    100.0 \%     &    100.0 \%   &  &   \\

   $      g_{Gunn}-[OIII]  $           &       0     &       1     &       2     &       2     &       2   &    0.13   &   0.07 \\
   $                       $          &      0.0 \%     &     50.0 \%     &    100.0 \%     &    100.0 \%     &    100.0 \%   &  &   \\

   $ g_{Gunn}-[OIII]/3000  $           &       1     &       2     &       3     &       4     &       4   &    0.40   &   0.19 \\
   $                       $          &     25.0 \%     &     50.0 \%     &     75.0 \%     &    100.0 \%     &    100.0 \%   &  &   \\

   $ g_{Gunn}-[OIII]/6000  $           &       0     &       0     &       0     &       1     &       1   &    0.50   &   0.23 \\
   $                       $          &      0.0 \%     &      0.0 \%     &      0.0 \%     &    100.0 \%     &    100.0 \%   &  &   \\
\hline
   $         Total         $           &      1    &      7    &     10    &     13    &     13    &    & \\
   $                       $          &      7.7 \%    &     53.8 \%    &     76.9 \%    &    100.0 \%    &    100.0 \%         \\
\enddata

\end{deluxetable}

\clearpage


\begin{deluxetable}{ccccccc}
\tabletypesize{\scriptsize} \tablecaption{Statistics by fields
with unresolved companions \label{statfield_qsos}}
\tablewidth{0pt} \tablehead{ \colhead{Filter Set} &
\multicolumn{5}{c}{Number of fields with companions with ...} &
\colhead{Number of fields}
\\
\colhead{} &
\colhead{$w<21.5$} &
\colhead{$w<22.5$} &
\colhead{$w<23.5$} &
\colhead{$w<24.5$} &
\colhead{any w} &
\\
}

\startdata
   $    F_{485}-HeII/3000  $          &     0   &     4   &     4   &     4   &     4    &  16  \\
   $                       $          &      0.0 \%     &     25.0 \%     &     25.0 \%     &     25.0 \%     &     25.0 \%      \\
   $    F_{485}-HeII/6500  $          &     0   &     0   &     0   &     1   &     1    &  4  \\
   $                       $          &      0.0 \%     &      0.0 \%     &      0.0 \%     &     25.0 \%     &     25.0 \%      \\
   $      g_{Gunn}-[OIII]  $          &     0   &     1   &     2   &     2   &     2    &  15  \\
   $                       $          &      0.0 \%     &      6.7 \%     &     13.3 \%     &     13.3 \%     &     13.3 \%      \\
   $ g_{Gunn}-[OIII]/3000  $          &     1   &     2   &     2   &     3   &     3    &  10  \\
   $                       $          &     10.0 \%     &     20.0 \%     &     20.0 \%     &     30.0 \%     &     30.0 \%      \\
   $ g_{Gunn}-[OIII]/6000  $          &     0   &     0   &     0   &     1   &     1    &  2  \\
   $                       $          &      0.0 \%     &      0.0 \%     &      0.0 \%     &     50.0 \%     &     50.0 \%      \\
\hline
   $          Total        $           &      1     &      7     &      8     &     11     &     11     &  47  \\
   $                       $              &    2.1 \%    &   14.9 \%    &   17.0 \%    &   23.4 \%    &   23.4 \%   & 100 \%     \\
\enddata

\end{deluxetable}

\clearpage


\begin{deluxetable}{cccccccc}
\tabletypesize{\scriptsize} \tablecaption{Statistics by number of
resolved companions \label{stat_res}} \tablewidth{0pt} \tablehead{
\colhead{Filter Set} & \colhead{$N(w<21.5)$} &
\colhead{$N(w<22.5)$} & \colhead{$N(w<23.5)$} &
\colhead{$N(w<24.5)$} & \colhead{$N(Total)$} & \colhead{\(
\frac{N}{N_{fields}} \)} & \colhead{\( \frac{N}{d\Omega/dz} ~
(\#/\square') \)}
\\
}

\startdata

   $    F_{485}-HeII/3000  $           &       4     &       9     &      17     &      24     &      35   &    2.19   &   1.05 \\
   $                       $          &     11.4 \%     &     25.7 \%     &     48.6 \%     &     68.6 \%     &    100.0 \%   &  &   \\
   $    F_{485}-HeII/6500  $           &       0     &       3     &       5     &       6     &      12   &    3.00   &   1.23 \\
   $                       $          &      0.0 \%     &     25.0 \%     &     41.7 \%     &     50.0 \%     &    100.0 \%   &  &   \\
   $      g_{Gunn}-[OIII]  $           &       3     &       7     &      12     &      15     &      18   &    1.20   &   0.59 \\
   $                       $          &     16.7 \%     &     38.9 \%     &     66.7 \%     &     83.3 \%     &    100.0 \%   &  &   \\
   $ g_{Gunn}-[OIII]/3000  $           &       4     &      14     &      18     &      21     &      24   &    2.40   &   1.14 \\
   $                       $          &     16.7 \%     &     58.3 \%     &     75.0 \%     &     87.5 \%     &    100.0 \%   &  &   \\
   $ g_{Gunn}-[OIII]/6000  $           &       2     &       2     &       4     &       5     &       7   &    3.50   &   1.60 \\
   $                       $          &     28.6 \%     &     28.6 \%     &     57.1 \%     &     71.4 \%     &    100.0 \%   &  &   \\
\hline
   $         Total         $           &     13    &     35    &     56    &     71    &     96    &    & \\
   $                       $          &     13.5 \%    &     36.5 \%    &     58.3 \%    &     74.0 \%    &    100.0 \%         \\
\enddata

\end{deluxetable}

\clearpage


\begin{deluxetable}{ccccccc}
\tabletypesize{\scriptsize} \tablecaption{Statistics by fields
with resolved companions \label{statfieldres}} \tablewidth{0pt}
\tablehead{ \colhead{Filter Set} & \multicolumn{5}{c}{Number of
fields with companions with ...} & \colhead{Number of fields}
\\
\colhead{} &
\colhead{$w<21.5$} &
\colhead{$w<22.5$} &
\colhead{$w<23.5$} &
\colhead{$w<24.5$} &
\colhead{any w} &
\\
}
\startdata
  $    F_{485}-HeII/3000  $          &     3   &     4   &     8   &    10   &    10    &  16   \\
   $                       $          &     18.8 \%     &     25.0 \%     &     50.0 \%     &     62.5 \%     &     62.5 \%        \\
   $    F_{485}-HeII/6500  $          &     0   &     3   &     3   &     3   &     4    &  4   \\
   $                       $          &      0.0 \%     &     75.0 \%     &     75.0 \%     &     75.0 \%     &    100.0 \%        \\
   $      g_{Gunn}-[OIII]  $          &     3   &     6   &     9   &     9   &    10    &  15   \\
   $                       $          &     20.0 \%     &     40.0 \%     &     60.0 \%     &     60.0 \%     &     66.7 \%        \\
   $ g_{Gunn}-[OIII]/3000  $          &     4   &     8   &     8   &     8   &     8    &  10   \\
   $                       $          &     40.0 \%     &     80.0 \%     &     80.0 \%     &     80.0 \%     &     80.0 \%        \\
   $ g_{Gunn}-[OIII]/6000  $          &     1   &     1   &     2   &     2   &     2    &  2   \\
   $                       $          &     50.0 \%     &     50.0 \%     &    100.0 \%     &    100.0 \%     &    100.0 \%        \\
\hline
   $          Total        $           &     11     &     22     &     30     &     32     &     34     &  47   \\
   $                       $              &   23.4 \%    &   46.8 \%    &   63.8 \%    &   68.1 \%    &   72.3 \%   & 100 \%      \\

\enddata
\end{deluxetable}

\begin{deluxetable}{lcccl}
\tabletypesize{\scriptsize} \tablecaption{QSO Candidates
Classification \label{spec-class}} \tablewidth{0pt} \tablehead{
\colhead{Candidate}& \colhead{$R.A.(2000)$} &
\colhead{$Dec.(2000)$} & \colhead{$z$} & \colhead{Lines; class}}
\startdata
cnd04     & 00 47 21.40 & -30 14 53.5  & 2.041  &   CIV,CIII]\\
cnd10     & 01 06 04.56 & -37 25 31.7  & 2.673  &   Ly$\alpha$,CIV,HeII\\
cnd14-1   & 01 28 18.11 & -35 18 50.5  & 1.466  &   CIV,CIII],MgII\\
cnd14-2   & 01 28 12.96 & -35 20 53.7  & 1.468  &   CIV,CIII],MgII\\
cnd18     & 03 18 18.39 & -20 09 40.6  & 2.891  &   Ly$\alpha$,CIV,HeII,CIII]; SiIV absorbed by CIV BAL?\\
cnd25     & 22 05 49.12 & -34 26 04.1  & 2.801  &   Ly$\alpha$, SiII, SiIV+OIV], CIV; BALQSO\\
cnd30     & 00 47 15.22 & -27 03 40.6  & ?      &   \\
cnd43-1   & 01 46 01.42 & -01 20 43.0  & 0.351  &   [OII],[OIII]5007; Starforming galaxy?\\
cnd43-2   & 01 45 49.17 & -01 21 10.5  & 3.136  &   OVI,Ly$\alpha$,SiIV,CIV\\
cnd44     & 01 46 11.59 & -00 43 27.9  & 3.066  &   Ly$\alpha$,CIV,CIII]\\
cnd52     & 04 22 15.98 & -38 42 08.3  & 1.614  &   CIV(w/absorption),CIII],MgII(weak)\\
cnd63     & 23 53 37.22 & -01 17 45.5  & 0.337  &   [OII],H$\gamma$,H$\beta$,[OIII]; Seyfert 2?\\
\enddata
\end{deluxetable}

\clearpage

\begin{thebibliography}{}
\bibitem[Adelberger \& Steidel(2000)]{Adelberger2000}
Adelberger, K.~L.~\& Steidel, C.~C.\ 2000, \apj, 544, 218
\bibitem[Bertin (1996)]{bertin96} Bertin, E. and Arnouts, S. 1996,
A\&A
\bibitem[Campos \etal (1999)]{campos99} Campos \etal ApJ Lett 511, L1
\bibitem[Croom \etal (2001)]{croom2001} Croom, S.~M., Shanks, T.,
Boyle, B.~J., Smith, R.~J., Miller, L., Loaring, N.~S., \& Hoyle, F.\ 2001, \mnras,
325, 483
\bibitem[Fynbo, M{\" o}ller, \& Thomsen(2001)]{fynbo2001}
Fynbo, J.~U., M{\" o}ller, P., \& Thomsen, B.\ 2001, \aap, 374, 443
\bibitem[Gunn \& Stryker(1983)]{gunn83} Gunn, J.~E.~\& Stryker, L.~L.\ 1983, \apjs, 52,
121
 1994, MNRAS 269, 277
\bibitem[Jacoby, Hunter, \& Christian(1984)]{jacoby84}
Jacoby, G.~H., Hunter, D.~A., \& Christian, C.~A.\ 1984, \apjs, 56, 257
\bibitem[Kauffmann \& Haehnelt(2000)]{kauffmann2000} Kauffmann, G.~\& Haehnelt, M.\ 2000, \mnras, 311, 576.
\bibitem[Keel \etal(1999)]{keel99}
Keel, W.~C., Cohen, S.~H., Windhorst, R.~A., \& Waddington, I.\ 1999, \aj, 118, 2547
\bibitem[Komberg \& Lukash (1994)]{komberg94} Komberg, B. V., \& Lukash, V.N.
\bibitem[Kurk et al. (2000)]{kurk2000} Kurk, J. D., R\"ottgering, H. J. A.,
Pentericci, L., Miley, G. K., van Breugel, W., Carilli, C. L., Ford, H.,
Heckman, T., McCarthy, P., \& Moorwood, A. 2000, A\&A, 358, L1
\bibitem[La Franca, Andreani, \& Cristiani(1998)]{franca98}
 La Franca, F., Andreani, P., \& Cristiani, S.\ 1998, \apj, 497, 529
\bibitem[M{\" o}ller \& Fynbo(2001)]{moller2001} M{\" o}ller, P.~\& Fynbo, J.~U.\ 2001,
\aap, 372, L57
\bibitem[del Olmo \& Moles(1991)]{olmo91} del Olmo, A.~\& Moles, M.\ 1991, \aap, 245,
27
\bibitem[Pei (1995)]{pei95} Pei, Y.C.\ 1995 \apj 1995, 438, 623.
\bibitem[Pentericci \etal (2000)]{pentericci2000} Pentericci, L., Kurk, J. D.,
R\"ottgering, H. J. A.,  Miley, G. K., van Breugel, W.,
Carilli, C. L., Ford, H., Heckman, T., McCarthy, P., \& Moorwood, A.
2000, A$\&A$ 341, 329
\bibitem[Rees (1984)]{rees84} Rees, M., 1984, ARAA 22, 471
\bibitem[Shanks and Boyle (1994)]{shanks94} Shanks, A. and Boyle,
1994,  MNRAS 271, 753
\bibitem[Stephens \etal (1997)]{stephens97} Stephens, A., Schneider, D.P.,
Schmidt, M., Gunn, J.E., \& Weimberg, D.H.\ 1997, \aj 511, 114, 41
\bibitem[Steidel \etal (2000)]{steidel2000} Steidel, C.~C.,
Adelberger, K.~L., Shapley, A.~E., Pettini, M., Dickinson, M.,
\& Giavalisco, M.\ 2000, \apj, 532, 170
\bibitem[Terlevich et al.(1991)]{T91} Terlevich, R.,
Melnick, J., Masegosa, J., Moles, M., \& Copetti, M.~V.~F.\ 1991, \aaps,
91, 285.
\bibitem[Tyson (1986)]{tyson86} Tyson, A. 1986, AJ 92, 691

\end{thebibliography}
\end{document}